\documentclass[aps,preprint,amsmath,amssymb, nofootinbib]{revtex4}
\usepackage{graphicx,color}
\def\be{\begin{eqnarray}}
\def\ed{\end{eqnarray}}
\def\non{\nonumber}
\def\ga{\gamma}

\def\la{\langle}
\def\ra{\rangle}

\begin{document}

\title{\Large \bf Rare B decays and  Tevatron top-pair asymmetry  }

\date{\today}

\author{\bf
Chuan-Hung~Chen$^{1,2}$\footnote{E-mail: physchen@mail.ncku.edu.tw
}, Sandy~S.~C.~Law$^{3}$\footnote{E-mail: slaw@cycu.edu.tw}, Run-Hui Li$^{4}$\footnote{E-mail: lirh@cskim.yonsei.ac.kr}}
\affiliation{
$^{1}$Department of Physics, National Cheng-Kung University, Tainan 701, Taiwan \\
$^{2}$National Center for Theoretical Sciences, Hsinchu 300, Taiwan
\\
$^{3}$Department of Physics, Chung Yuan Christian University, Chung-Li 320, Taiwan \\
$^{4}$Department of Physics $\&$ IPAP, Yonsei University, Seoul 120-479, Korea
}

\begin{abstract}
The recent Tevatron result on the top quark forward-backward
asymmetry, which deviates from its standard model prediction by
3.4$\sigma$, has prompted many authors to build new models to
account for this anomaly. Among the various proposals, we find that
those mechanisms which produce $t\bar t$ via $t$- or $u$-channel can
have a strong correlation to the rare $B$ decays. We demonstrate
this link by studying a model with a new charged gauge boson, $W'$.
In terms of the current measurements on $B\to \pi K$ decays, we
conclude that the branching ratio for $B^-\to \pi^- \bar K^0$ is
affected most by the new effects. Furthermore, using the world
average branching ratio for the exclusive $B$ decays at $2\sigma$
level, we discuss the allowed values for the new parameters.
Finally, we point out that the influence of the new physics effects
on the direct CP asymmetry in $B$ decays is insignificant.
\end{abstract}

\maketitle

Recently, the forward-backward asymmetry (FBA) in top-pair production  is measured  by  the D{\O} \cite{D0_PRL100} and CDF \cite{CDF_PRL101} Collaborations  in $p\bar p$ collisions at $\sqrt{s}=1.96$ TeV. With an integrated luminosity of $5.3$ fb$^{-1}$, CDF further finds that  the large top quark FBA  occurs at large $t\bar t$ rapidity difference ($\Delta y$) and invariant mass of $t\bar t$ ($M_{t\bar t}$) \cite{Aaltonen:2011kc}:
 \be
 A^{t\bar t}(|y| < 1.0) &=& 0.026 \pm 0.118 \, [ 0.039 \pm 0.006] \,, \non \\
 A^{t\bar t}(|y| \geq 1.0) &=& 0.611 \pm 0.256\, [ 0.123 \pm 0.008 ]\,, \non \\
 A^{t\bar t}(M_{t\bar t} < 450 \textrm{ GeV}) &=& -0.116 \pm 0.153\,  [0.040 \pm 0.006]\,, \non \\
  A^{t\bar t}(M_{t\bar t} \geq 450 \textrm{ GeV}) &=& 0.475 \pm 0.114\, [0.088 \pm 0.013]\,,
  \label{equ:data}
 \ed
where the value in the square brackets  denotes the theoretical
result calculated by Monte Carlo program (MCFM)
\cite{Campbell:1999ah}  at next-to-leading order (NLO). The
inconsistency  between data and standard model (SM) predictions
displayed in Eq.~(\ref{equ:data}) indicates that new physics effects
may be at play and hence leading to this anomalous result.

\begin{figure}[hptb]
\includegraphics*[width=3.5 in]{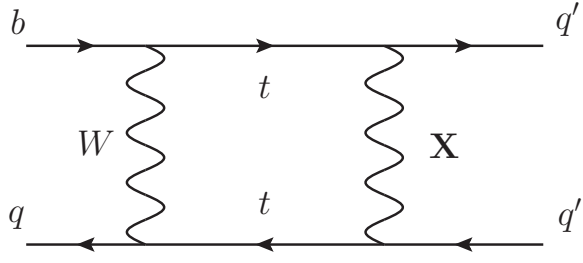}
\caption{   Feynman diagrams for $b\to q' \bar q' q$ with $q'=d$ or $u$ and $q=d, s$, where ${\bf X}$ stands for a generic new particle.}
 \label{fig:Feyn}
\end{figure}

In the wake of this $3.4\sigma$ deviation of the  observed value of
the top quark FBA from the SM predicted one, several possible
solutions have been proposed and studied by authors in
Refs.~\cite{AKR,Djouadi:2009nb,Ferrario:2009bz,Jung:2009jz,Cheung:2009ch,Frampton:2009rk,Shu:2009xf,Arhrib:2009hu,Ferrario:2009ee,Dorsner:2009mq,Jung:2009pi,Cao:2009uz,Barger:2010mw,Cao:2010zb,Kumar:2010vx,Martynov:2010ed,Chivukula:2010fk,Bauer:2010iq,Chen:2010hm,Jung:2010yn,Alvarez:2010js,Choudhury:2010cd,Cheung:2011qa,Xiao:2011kp,Delaunay:2011vv,Berger:2011ua,Barger:2011ih,Bhattacherjee:2011nr,Blum:2011up,Patel:2011eh,Isidori:2011dp,Foot:2011xu,Delaunay:2011gv,AguilarSaavedra:2011vw,Gresham:2011pa,Shu:2011au}.
Among these new mechanisms, we see that a potentially interesting
correlation to $B$ meson physics may arise. In particular, the new
interaction that contributes to $t\bar t$ production by $t$- or
$u$-channel will also contribute to rare $B$ decays, e.g. $B\to \pi
(\pi, K) $. This is achieved through box diagrams like those
sketched in Fig.~\ref{fig:Feyn}, where ${\bf X}$ could be a colored
vector \cite{AKR,Ferrario:2009bz,Frampton:2009rk}, $Z'$
\cite{Jung:2009jz}, $W'$
\cite{Cheung:2009ch,Barger:2010mw,Cheung:2011qa,Barger:2011ih} or
colored scalar bosons
\cite{Shu:2009xf,Arhrib:2009hu,Dorsner:2009mq,Patel:2011eh}, while
$q'=d$ or $u$ depending on the charge of the ${\bf X}$ particle. It
is known that with the enormous $B$ meson production at LHCb, BaBar,
Belle and Tevatron, the errors in the measured branching ratios
(BRs) for $B\to \pi K$ decays have now reached percent level. As a
result of this high accuracy, it is interesting to investigate how
strong is the correlation between rare $B$ decays and top quark FBA.
In addition, one can study the influence on the associated CP
asymmetry (CPA) in such new physics setups. We present our analyzes
on both issues in the following. 

\begin{figure}[hptb]
\includegraphics*[width=4.5 in]{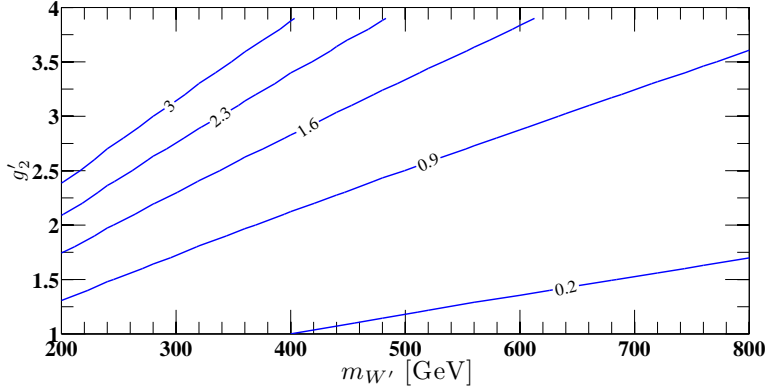}
\caption{  Contours for $C_{W'}$ as a function of $g'_2$ and $m_{W'}$, where the number on each curve denotes the value of $C_{W'}$ in units of $10^{-2}$.}
 \label{fig:cWp}
\end{figure}

In order to illustrate the impact of the new physics (which leads to
a large top-pair FBA as measured) on the low energy $B$ sector, in
this paper, we shall focus on the case with a new $W'$ interaction,
namely, the $t$-$d$-$W'$ coupling.  A similar discussion could be
applied to other new interactions in the $Z'$, colored vector or
colored scalar models.
Since we are studying the impact of new physics on $B$ decays, the
detailed analysis of top-pair production could be referred to those
papers listed in the references. We start by writing the relevant
interaction as \cite{Barger:2010mw}
 \be
 {\cal L}=-i\frac{g'_2}{\sqrt{2}} \bar t \ga^\mu W'_\mu P_R d  + h.c\,, \label{eq:intWp}
 \ed
where $g'_2$ is the new gauge coupling which is regarded as a free parameter.  By including the contributions of $W$ and  the associated Goldstone bosons, the Hamiltonian for $b\to q d \bar d$ ($q= d, s$) shown in Fig.~\ref{fig:Feyn} is  given by
 \be
 {\cal H}_{W'} &=& -\frac{G_F}{\sqrt{2}} V_{tb} V^*_{tq} \left( \frac{g'_2 m_t}{4\pi m_{W'}}\right)^2 I(x_W,x_t)\bar d (1-\ga_5) b \bar q(1+\ga_5) d \,, \non \\
&=&  - \frac{G_F}{\sqrt{2}} V_{tb} V^*_{tq} C_{W' } \bar d (1-\ga_5) b \bar q(1+\ga_5) d \label{eq:int_H}
 \ed
where $C_{W'}$ stands for the new Wilson coefficient at electroweak scale, $x_W=m^2_W/m^2_{W'}$, $x_t=m^2_t/m^2_{W'}$ and  $I(x_W,x_t)= (1+x_W)I_1(x_W,x_t) + 2 (x_W + x_t) I_2(x_W,x_t)$ with
 \be
 I_1(a,b) &=& \int^1_0 dz_1 \int^{z_1}_0 dz_2 \frac{z_2}{ 1-(1-a)z_1 -(a-b) z_2}\,, \non \\
 I_2(a,b)&=& \int^1_0 dz_1 \int^{z_1}_0 dz_2 \frac{z_2}{ (1-(1-a)z_1 -(a-b) z_2)^2}\,.
 \ed
  We note that there are only two new free parameters in $C_{W'}$. To illustrate the dependence on $g'_2$ and the mass of $W'$, we plot the contours for $C_{W'}$ as a function of $g'_2$ and $m_{W'}$ in Fig.~\ref{fig:cWp}. The number on each curve in Fig.~\ref{fig:cWp} denotes the value of $C_{W'}$ in units of $10^{-2}$. For instance, with $g'_2 =3$ and $m_{W'}=700$~GeV, we get $C_{W'}=0.85\times 10^{-2}$.

  Notice that if we perform Fierz transformation on the fields in Eq.~(\ref{eq:int_H}), the four-Fermi operator could be transformed into $\bar d_\alpha  \ga_\mu (1+\ga_5) d_\beta \bar q_\beta \ga^\mu (1-\ga_5) b_\alpha$, where  $\alpha, \, \beta$ denote the color indices. This operator is the same as the  $O_6$ operator in the SM \cite{Buchalla:1995vs} and its corresponding Wilson coefficient at $m_W$ scale is $C_6 \simeq -0.2\times 10^{-2}$. According to our brief analysis, we see that the new interaction has the same behavior as $O_6$, and so, its contribution should be similar to that dictated by $C_6$.

\begin{figure}[hptb]
\includegraphics*[width=5 in]{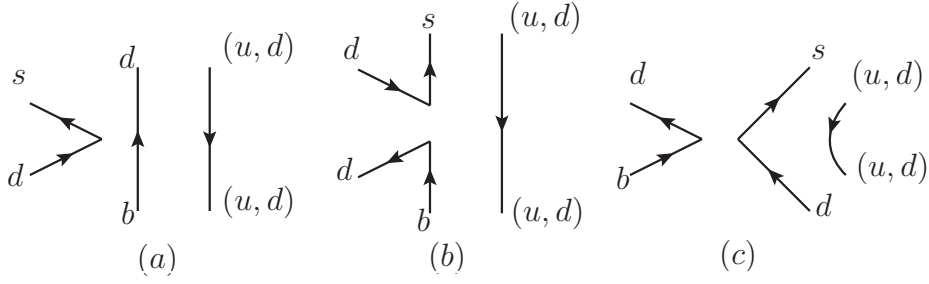}
\caption{Flavor diagrams for $B\to \pi K$ decays: (a) color-allowed emission, (b) color-suppressed emission, and (c) annihilation diagrams.}
 \label{fig:FD}
\end{figure}

Since  the new four-Fermi interactions in Eq.~(\ref{eq:int_H}) are induced by loop diagrams, we  expect that the influence of the new effects on the tree dominant $B\to \pi \pi $  decays should be small. Therefore, we only concentrate on the $B\to \pi K$ decays. This is because at quark level, they are associated with the $b\to s d \bar d$ process which is dominated by gluonic penguins in the SM. In order to comprehend how the new effects contribute to each decay mode in the $B\to \pi K$ processes, we draw the flavor diagrams in Fig.~\ref{fig:FD}, where figures (a) and (b) denote the color-allowed and -suppressed emission diagrams respectively, while figure (c) represents the annihilation diagram. Importantly, we note that the decay  $B^-\to \pi^- \bar K^0$,  $B^- \to \pi^0 K^-$ and $\bar B_d \to \pi^+ K^-$ are influenced by Fig.~\ref{fig:FD}(a), (b) and (c) respectively,  whereas  $\bar B_d \to \pi^0 \bar K^0$ involves all three figures.

It is well-known that the calculations of a nonleptonic exclusive decay  include  factorizable and nonfactorizable parts. Owing to the color flow, the latter is always color-suppressed and smaller than the former. Hence, to simplify the analysis, we only calculate the factorizable contributions from the $W'$ exchange when displaying the influence of the new physics effects.  However, we keep both parts  when accounting for  the  SM sector.  In order to study the nonleptonic exclusive $B$ decays, we parametrize the relevant decay constants, transition and time-like form factors as
 \be
 \la 0 | \bar q_2 \ga^\mu \ga_5 q_1 | \bar M\ra &=& if_M P^\mu \,, \non \\
 \la \bar M | \bar q_1 \ga^\mu \ga_5 q_2 |0 \ra &=& -i f_M P^\mu \,, \non \\
 \la \bar  M' | \bar q_2 \ga^\mu q_1 | \bar M\ra &=& f^{MM'}_{+}(q^2) \left( P^\mu - \frac{P\cdot q}{q^2} q^\mu \right) + f^{MM'}_{0}(q^2) \frac{P\cdot q}{q^2} q^\mu \,, \non \\
 \la \bar M_1 M_2 | \bar q_1 \ga^\mu q_2 | 0 \ra &=& g^{M_1 M_2}_{+}(Q^2) \left( \tilde{q}^\mu - \frac{Q\cdot \tilde{q} }{Q^2}Q^\mu  \right)+ g^{M_1 M_2}_0 (Q^2) \frac{Q\cdot \tilde{q} }{Q^2}Q^\mu  \label{eq:ff1}
 \ed
with $P=P_{M} + P_{M'}$, $q=P_M-P_{M'}$, $Q=P_{M_1}+P_{M_2}$ and $\tilde{q}= P_{M_1}-P_{M_2}$. Applying the equations of motion, we then get
 \be
  \la 0 | \bar q_2 \ga_5 q_1 |\bar M\ra &=&  \la \bar M |\bar q_1 \ga_5 q_2 | 0 \ra = -i f_M \frac{m^2_M}{m_{q_1} + m_{q_2}}\,, \non \\
 \la \bar M' | \bar q_2  q_1 |\bar M\ra &=&\frac{M^2 -M'^2}{m_{q_1} - m_{q_2} }  f^{MM'}_{0}(q^2) \,, \non \\
  \la \bar M_1 M_2 | \bar q_1  q_2 |0 \ra &=& \frac{M^2_1 - M^2_2}{m_{q_1} - m_{q_2} } g^{M_1 M_2}_0 (Q^2) \,. \label{eq:ff2}
 \ed
Using the above form factors, the decay amplitude with $W'$ exchange is formulated by
 \be
%
%
 %
 A^{W'}(\pi K) =  \frac{G_F V_{tb} V^*_{ts}}{\sqrt{2}} C_{W'}(\mu_b) \zeta_{\pi K}
 \label{equ:amp1}
  \ed
for $ B^- \to \pi^- \bar K^0$, $\bar B_d\to \pi^+ K^-$ and $B^-\to
\pi^0 K^-$  processes,
  while
   \be
 A^{W'} (\pi^0 \bar K^0 )&=& -\frac{1}{\sqrt{2}} A^{W'} ( \pi^- \bar K^0) + A^{W'}(\pi^0 \bar K^-) - \frac{1}{\sqrt{2}} A^{W'}(\pi^+ K^-)
   \ed
 for $\bar B_d\to \pi^0 \bar K^0$,  where
  \be
 \zeta_{\pi^- \bar K^0} &=&  f_K \frac{m^2_K}{m_s + m_d} \frac{m^2_B - m^2_\pi }{m_b -m_u } f^{B\pi}_0(m^2_K)\,, \non \\
 \zeta_{\pi^+  K^-} &=& f_B \frac{m^2_B}{m_b + m_u} \frac{m^2_K - m^2_\pi }{m_s -m_d } g^{K\pi}_0(m^2_B)  \,, \non \\
\zeta_{\pi^0 K^-}   &=&  f_\pi \frac{ m^2_B -m^2_K }{2\sqrt{2} N_c} f^{BK}_{0} (m^2_\pi)    \,.
  \ed
 Clearly, once the first three decays have been determined, the last one will be fixed. Since $m^2_{K, \pi}\ll m^2_B$, it is a good approximation to take $f^{B\pi(K)}_0 (m^2_{K(\pi)}) \approx f^{B\pi(K)}_0 (0)$. Moreover, the transition form factor $f^{BM}_0(0)$ and time-like form factor $g^{M_1 M_2}_0(m^2_B)$ are associated with SM QCD, and for calculating them, we employ the perturbative QCD approach (PQCD)  \cite{KpPQCD}. The  resulting formulae are summarized in Appendix~\ref{sec:app}. We note that in Eq.(\ref{equ:amp1}), $C_{W'}(\mu_b)$ is the new Wilson coefficient at $\mu_b = \mathcal{O}(m_b)$ scale.

Now that the necessary formulations have been built up, we list the  SM theoretical inputs required for our subsequent analysis in Table~\ref{tab:hardfun}, where $V_{ub}=|V_{ub}|e^{-i\phi_3}$. Using these values and formulations in the Appendix, the transition and time-like form factor are found to be
 \be
 f^{B\pi}_0 (0)&\approx & 0.24\,, \non \\
  f^{BK}_0 (0) &\approx & 0.35 \,, \non \\
  g^{\pi K}_0 (m^2_B) &\approx & 0.15e^{i 1.1} \,.
 \ed
For estimating $C_{W'}(\mu_b)$, we adopt the relation: $C_{W'}(\mu_b) \approx \left(\alpha_s(\mu_b)/\alpha_s(m_W)\right)^{\tilde d}  C_{W'}(m_W)$ with $\tilde{d} =N_c(N^2_c-1)/(11N_c -2 n_f)$, $N_c=3$ and $n_f$ being the number of effective quark flavors \cite{YS1975}.
\begin{table}[hptb]
\caption{The values of  theoretical inputs. } \label{tab:hardfun}
\begin{ruledtabular}
\begin{tabular}{ccccc}
$|V_{ub}|$ & $\phi_3 $ (deg) & $V_{ts}$ & $ m^0_\pi $ [GeV] & $m^0_K $ [GeV]  \\ \hline
$3.9\times10^{-3}$ & $65$ & $-0.041$ & 1.34 & 1.70  \\  \hline
$m_{d(u)}$ [MeV] & $m_s$ [MeV] &$f_\pi$ [MeV] & $f_K$ [MeV] & $f_B$ [MeV]  \\ \hline
4.5 & 100 & 130 & 160 & 190  \\
   \end{tabular}
\end{ruledtabular}
\end{table}
In order to combine the SM results with the effects of $W'$ exchange, we write the total decay amplitude for  $B\to \pi  K$ as
 \be
 A(\pi K) &=& A^{SM}(\pi K) + A^{W'}(\pi K)\,.
 \ed
As usual, the SM contributions could be expressed by \cite{Chen:2008ie}
\be
 A^{SM}(\pi^- \bar K^0) &=& -\frac{ G_F V_{tb} V^*_{ts}}{\sqrt{2} }  P \,, \non \\
 A^{SM}(\pi^+ K^-) &=&  - \frac{ G_F V_{tb} V^*_{ts}}{\sqrt{2} }  P - \frac{ G_F V_{ub} V^*_{us}}{\sqrt{2} }  T \,, \non \\
 \sqrt{2} A^{SM}(\pi^0 K^-) &=&- \frac{ G_F V_{tb} V^*_{ts}}{\sqrt{2} }  (P + P_{EW} )- \frac{ G_F V_{ub} V^*_{us}}{\sqrt{2} }  (T +C) \,, \non \\
 \sqrt{2}A^{SM}(\pi^0 K^0) &=&  \frac{ G_F V_{tb} V^*_{ts}}{\sqrt{2} }  (P -P_{EW}) - \frac{ G_F V_{ub} V^*_{us}}{\sqrt{2} }  C \,. \label{equ:notat41}
 \ed
In Eq.~(\ref{equ:notat41}), $P$, $P_{EW}$, $T$ and $C$ denote, respectively, the contributions from gluonic penguin, electroweak penguin, color-allowed tree and color-suppressed tree. To deal with these hadronic effects, we quote the results calculated by PQCD, where the values in the SM are given by  \cite{Li:2005kt}
 \be
P &=& 0.15 e^{-i0.24} \,, \ \ \ P_{EW} = 0.018 e^{i 0.44}\,, \non \\
T &=& 1.05 e^{i 0.1}  \,, \ \ \ C= 0.270  e^{-i 1.3 } \,.
 \ed
Consequently, the resulting SM predicted values of the BRs and CPAs  are presented along with the experimental data \cite{ TheHeavyFlavorAveragingGroup:2010qj} in Table~\ref{tab:th_exp}.
\begin{table}[hptb]
\caption{SM results \cite{Li:2005kt} and experimental data \cite{TheHeavyFlavorAveragingGroup:2010qj} for BRs and CPAs of $B\to \pi K$ decays. } \label{tab:th_exp}
\begin{ruledtabular}
\begin{tabular}{ccccc}
Decay &$B^-\to \pi^- \bar K^0$ & $\bar B_d \to \pi^+ K^-$  & $B^-\to \pi^0 K^- $ & $ \bar B_d \to \pi^0 \bar K^0 $    \\ \hline
${\cal B}^{SM} [10^{-6}]$ & $23.5$ & $20.46$ & $ 13.25$ & $9.16$   \\
${\cal B}^{Exp} [10^{-6}]$ & $23.1\pm 1.0 $ & $19.4 \pm 0.6 $ & $ 12.9\pm 0.6 $ & $9.5 \pm 0.5 $   \\  \hline
$A^{SM}_{CP} [\%]$ & $ 0$ & $ -10.18$  & $-0.95$ & $-6.29$   \\
$A^{Exp}_{CP} [\%]$  & $0.9 \pm 2.5$ & $-9.8^{+1.2}_{-1.1}$ & $5 \pm 2.5$ & $-1\pm 10$  \\
   \end{tabular}
\end{ruledtabular}
\end{table}

\begin{figure}[tb]
\includegraphics*[width=5 in]{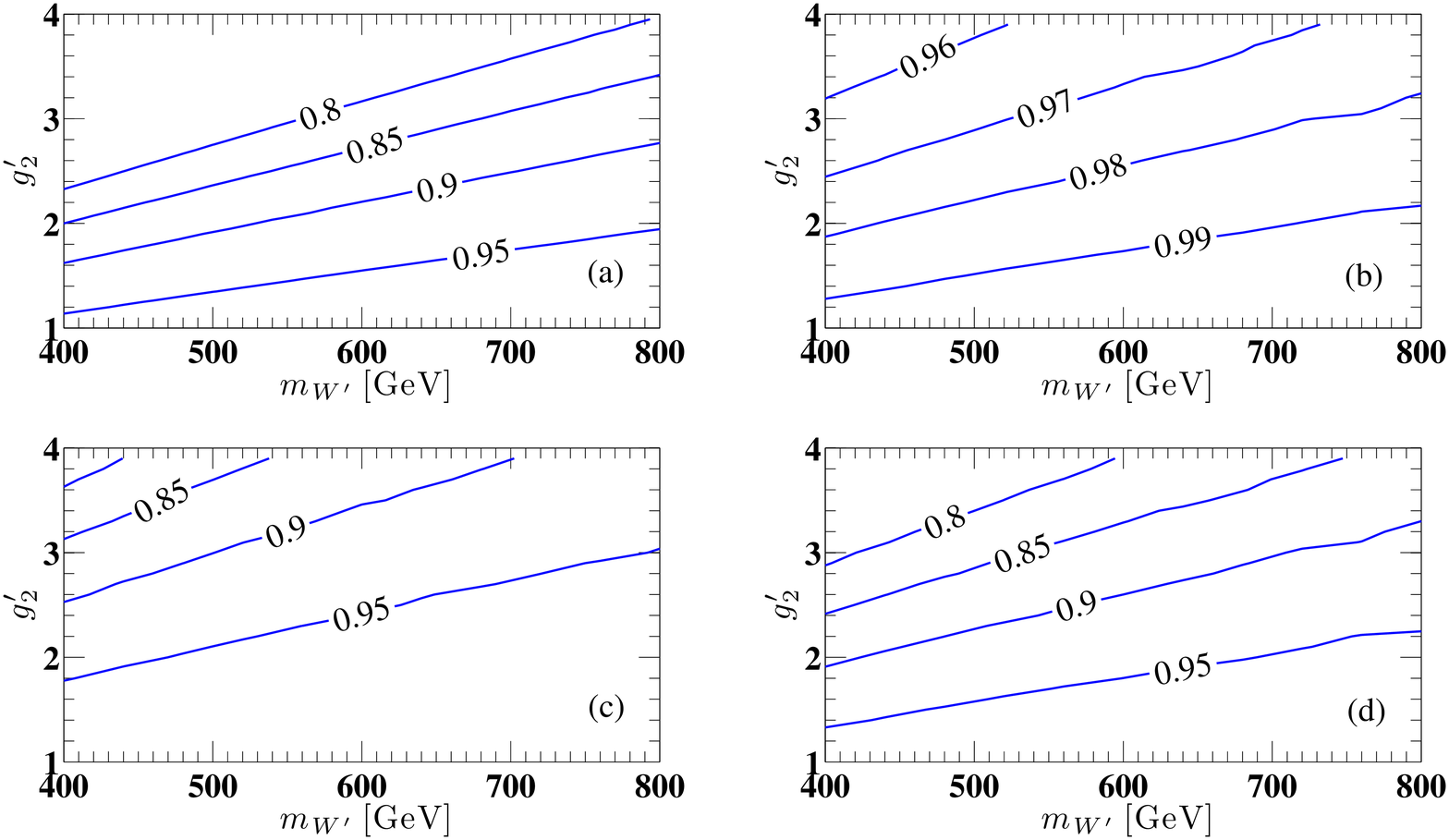}
\caption{ Contours for the ratio $R_B$ as a function of $g'_2$ and $m_{W'}$, where the correspondence of each plot is (a) $B^-\to \pi^- \bar K^0$, (b) $\bar B_d \to \pi^+ K^-$,  (c) $B^-\to \pi^0 K^-$ and (d) $\bar B_d\to  \pi^0 \bar K^0$.  }
 \label{fig:contBr}
\end{figure}

To display the influence of the $W'$ effects on BRs and CPAs in $B$ decays, we consider the two quantities $R_B$ and $R_{CP}$ which represent, respectively, the ratios of BR and CPA in the $W'$ exchange to those in the SM:
 \be
 R_B &\equiv& \frac{{\cal B}^{W'}(B\to \pi K)}{{\cal B}^{SM}(B\to \pi K)}\,, \non \\
 R_{CP} &\equiv&  \frac{A^{W'}_{CP} (B\to \pi K)}{A^{SM}_{CP}(B\to \pi K)}\,. \label{eq:ratios}
 \ed
Because there is  only two new free parameters in the $W'$-mediated
effects, we use contour plots as a function of $g'_2$ and $m_{W'}$
to display the deviation from the SM predictions. With the SM inputs
discussed earlier, the contours for $R_B$ as a function of $g'_2$
and $m_{W'}$ are shown in Fig.~\ref{fig:contBr}, where (a)-(d)
correspond to the decays $B^-\to \pi^- \bar K^0$, $\bar B_d \to
\pi^+ K^-$,  $B^-\to \pi^0 K^-$ and  $\bar B_d\to  \pi^0 \bar K^0$,
respectively. From this, we see that $W'$-mediated effects have a
significant contribution to the BRs of the $\pi^- \bar K^0$, $\pi^0
K^-$ and $\pi^0 \bar K^0$ modes, whereas the effects are small for
the $\pi^+ K^-$ mode.  The main reason for this can be traced to the
fact that the associated topology for $\bar B_d\to \pi^+ K^-$ decay
is of annihilation type,  whose corresponding hadronic effects are
usually smaller than those arising from the emission diagrams.

\begin{figure}[tb]
\includegraphics*[width=4.5 in]{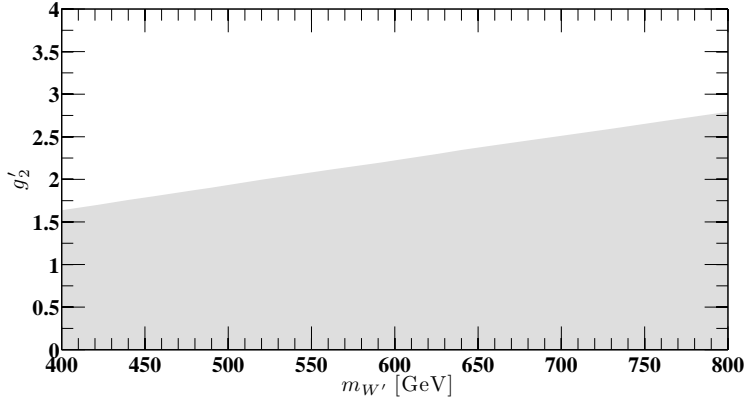}
\caption{  Allowed range (gray) for $g'_2$ and $m_{W'}$, where the world average BR for $B^-\to \pi^- \bar K^0$ with $2\sigma$ errors has been used. }
 \label{fig:g2_mWp}
\end{figure}

In addition, if we compare the three decay modes $\pi^- \bar K^0$,
$\pi^0 K^-$ and $\pi^0 \bar K^0$, one finds that the influence on
$\pi^0 K^-$ mode is smaller than the other two because
Fig.~\ref{fig:FD}(b) is color-suppressed. From our numerical
results, we conclude that $B^-\to \pi^- \bar K^0$ decay is the most
sensitive to the $W'$-mediated effects. In order to display the
constraint from the measured data, we plot the allowed range for
$g'_2$ and $m_{W'}$ in Fig.~\ref{fig:g2_mWp}, where we have adopted
the world average BR for $B^-\to \pi^- \bar K^0$ with $2\sigma$
errors, $(2.31\pm0.20)\times
10^{-5}$ \cite{Amsler:2008zzb}, and the PQCD results with the uncertainties are taken as $|P|=0.15^{+0.04}_{-0.03}$ and $arg(P)=-0.24^{+0.1}_{-0.2}$ \cite{Li:2005kt}. From these, we find that only those values of $|P|$ approaching to the central value calculated by PQCD contribute to the range $400 < m_{W'}<800$ and the allowed $m_{W'}$ from other values of $|P|$ is below $200$ GeV, which is disfavored by the analysis in Ref.~\cite{Barger:2010mw}. Based on the result, we observe that $g'_2
\gtrsim 2.5 $ is excluded when $m_{W'} \lesssim 700$ GeV. Such
constraint on the parameter space highlights the tension between the
need to choose a large enough coupling ($g_2' \simeq 3$ when $m_W'
\simeq 550$~GeV \cite{Barger:2011ih}) for this type of model to
explain the top anomaly and satisfying the limits implied by rare B
processes at the same time.\footnote{Note that the choice of the numerical values used here is for convenience and ease of comparison with existing work such as  \cite{Barger:2011ih}. This selection of parameter space would not in any way bias our conclusion.} As a result, we expect that new models
which have the $t \bar{t}$ produced via $t$- or $u$-channel are
subject to similar restrictions when fine-tuning for a workable
parameter set.

\begin{figure}[tb]
\includegraphics*[width=6 in]{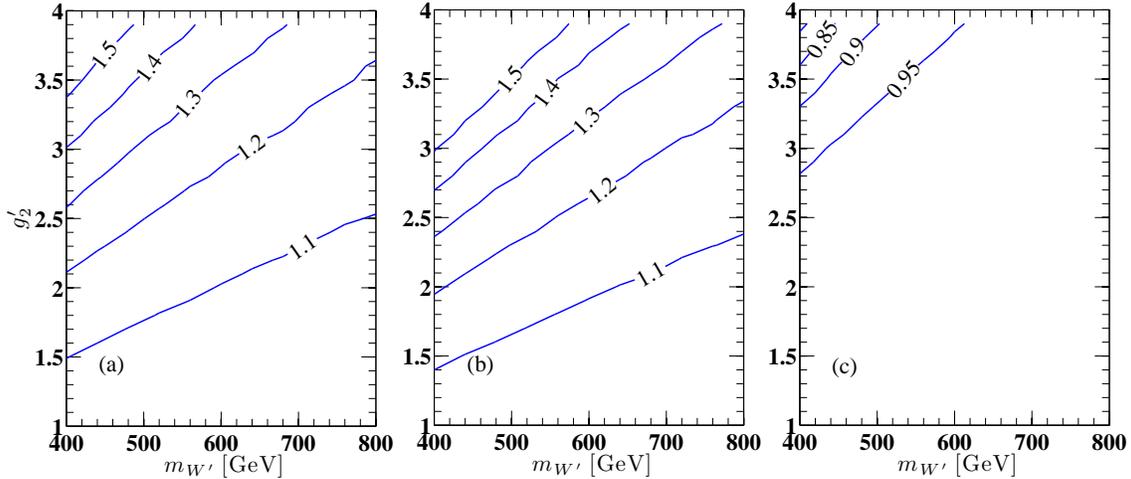}
\caption{  Ratio of CPA in the $W'$ exchange to that in the SM for (a) $\bar B_d\to \pi^+ K^-$, (b) $B^-\to \pi^0 K^-$ and (c) $\bar B_d\to \pi^0 \bar K^0$.}
 \label{fig:CPA}
\end{figure}

Similarly, using the definition in Eq.~(\ref{eq:ratios}), the ratio
$R_{CP}$ as a function of $g'_2$ and $m_{W'}$ is plotted in
Fig.~\ref{fig:CPA}, where (a)-(c) correspond to the decays $\bar
B_d\to \pi^+ K^-$, $B^-\to \pi^0 K^-$ and $\bar B_d\to \pi^0 \bar
K^0$, respectively. Looking at Fig.~\ref{fig:CPA}(b), it seems that
there is a sizable effect on the CPA for $B^-\to \pi^0 K^-$.
However, because the SM result for $A_{CP}(B^-\to \pi^0 K^-)$ is
small, the resulting shift in magnitude of CPA due to the presence
of new physics is only around $1\%$. Furthermore, if we combine the
result in Fig.~\ref{fig:CPA}(a) with the allowed range depicted in
Fig.~\ref{fig:g2_mWp}, we see that the modification from the $W'$
effects will be about $10\%$ of $A^{SM}_{CP}(\bar B_d\to \pi^+ K^-)$
only. Therefore, the change in the absolute value of CPA for $\bar
B_d \to \pi^+ K^-$ is at about the $1\%$ level. Finally, we note
from Fig.~\ref{fig:CPA}(c) that the CPA for $\bar B_d\to \pi^0 \bar
K^0$ is more or less insensitive to the $W'$ effects.
%

In summary, the fascinating discovery of an unexpectedly large FBA
in the top quark pair production at the Tevatron strongly suggests
that some new physics may be at play. While the exact nature of this
is not clear, we observe that any new interactions introduced to
solve the anomaly which produce $t\bar t$ via the $t$- or
$u$-channel are naturally correlated to rare $B$ decays. In this
work, we illustrate this connection by studying a model with a new
charged gauge boson $W'$ added to the SM. Using the current
precision measurements on $B\to \pi K$ decays, we estimate the
constraints on the parameters of the theory. In particular, we find
that strong conditions are imposed on the new gauge coupling and the
mass of the associated gauge boson.
 From the data for $B^-\to \pi^- \bar K^0$, which is the most stringent among the set of $B\to \pi K$ decays considered, our study shows that
the size of the coupling $g_2'$ is strongly constrained from above for the relevant $W'$ mass range of several hundreds GeV.
On the other hand, these $W'$-mediated effects are not expected to play a significant role on the direct CPA of $B$ decays.


\begin{acknowledgments}
\noindent  CHC is supported in part by the National Science
Council (NSC) of R.O.C. under Grant No. NSC-97-2112-M-006-001-MY3. SSCL is supported in part by the NSC of R.O.C. under Grant No. NSC-99-2811-M-033-013 and in part by the National Center for Theoretical Sciences (NCTS) of Taiwan. RHL is supported by the Brain Korea 21 project.

\end{acknowledgments}


\begin{appendix}
\section{formulations of $f_0^{BM}(0)$ and $g^{M_1M_2}_0(m^2_B)$  in PQCD approach}\label{sec:app}

For calculating the transition and time-like form factors in PQCD approach, the necessary
distribution amplitude  of $\bar B$ meson and outgoing light meson is defined by
\be
 \la 0| \bar{q}_{\beta}(z) b_{\alpha}(0) |\bar B(p_B)\ra &=&
 \frac{i}{\sqrt{2N_c}}\int d^4k e^{ik\cdot z}\left\{(\not p_B +m_B)\gamma_5\left[\phi_B(k)
 -\frac{\not n_+ -\not n_-}{\sqrt{2}}\bar\phi_B(k)\right]\right\}_{\alpha\beta} \,, \non \\
 \la \bar M(p) | \bar{q}_{2\beta}(z) q_{1\alpha}(0) |0\ra &=& -
\frac{i}{\sqrt{2N_{c}}} \int^{1}_{0} dx e^{ix p\cdot z} \left\{
\gamma_{5} \not p \Phi_{M} (x) + \gamma_{5} m^{0}_{M} \Phi^{p}_{M}
(x) \right. \non \\
&& \left. + m^{0}_{M} \gamma_{5} (\not n_{+} \not n_{-} -1)
\Phi^{\sigma}_{M} \right\}_{\alpha \beta}\,.
 \ed
with $p_B=\frac{m_B}{\sqrt{2}}(1,1,0_{\perp})$, $p=(p^{+},0, 0_{\perp})$, $n_{+}=(1,0,0_{\perp})$,  $n_{-}=(0,1,0_{\perp})$. Since the effects of $\bar\phi_B$ are small \cite{Lu:2002ny},  in the calculations we
only focus on the contributions from $\phi_B$. By hard gluon exchange, the $\bar B\to \bar M$ transition form factor is expressed by
 \be
f^{BM}_{0}(0)&=&
 8\pi C_{F} m_{B}^{2} \int_{0}^{1}
dx_{1}dx_{2}\int_{0}^{\infty }b_{1}db_{1}b_{2}db_{2}\Phi _{B}\left(
x_{1},b_{1}\right)
\nonumber \\
&&\times \left\{ \left[\left(1+x_{2}\right) \Phi _{M}\left(
x_{2}\right) +r_{M} ( 1-2x_{2}) \left( \Phi _{M}^{p}\left(
x_{2}\right) +\Phi _{M}^{\sigma }\left( x_{2}\right) \right)
\right]E_{e}\left( t_{e}^{\left( 1\right) }\right) \right.
\nonumber \\
&&\times \left. h_{e}\left( x_{1},x_{2},b_{1},b_{2}\right) +
2r_{M}\Phi _{M}^{p}\left( x_{2}\right)  E_{e}\left( t_{e}^{\left(
2\right) }\right) h_{e}\left( x_{2},x_{1},b_{2},b_{1}\right)
\right\}
 \ed
with  $C_F=4/3$ and $r_M=m^0_{M}/m_{B}$, and the $0\to \bar M_1 M_2$ time-like form factor is written as
 \be
g_0^{M_1M_2}(m^2_B)&=& -\frac{(m_b +m_u)(m_s-m_d)}{m^2_K-m^2_\pi }8\pi
C_{F}m_{B}^{2}\int_{0}^{1}dx_{2}dx_{3}\int_{0}^{\infty}
b_{2}db_{2}b_{3}db_{3}
\nonumber \\
&&\times \left\{ \left[  r_{M_2} x_{2} \Phi _{M_1}(x_{2}) \left(
\Phi _{M_2}^{p} ( 1-x_{3}) + \Phi_{M_2}^{\sigma }(1-x_{3}) \right)
+2r_{M_1 }\Phi^{p} _{M_1}( x_{2}) \Phi_{M_2} ( 1-x_{3}) \right]
\right.
\nonumber \\
&&\times E_{a}\left( t_{a}^{1 }\right) h_{a}(
x_{2},x_{3},b_{2},b_{3}) + \left[ x_{3} r_{M_1}
\left(\Phi_{M_1}^{p}( x_{2})-\Phi _{M_1}^{\sigma}( x_{2}) \right)
\Phi_{M_2}(
1-x_{3}) \right. \nonumber \\
&& \left. \left. + 2r_{M_2}\Phi_{M_1}(x_2)\Phi
_{M_2}^{p}(1-x_3)\right] E_{a}\left( t_{a}^{2 }\right)
h_{a}(x_{3},x_{2},b_{3},b_{2}) \right\}\,.
 \ed
The hard functions $h_{e, a}$ are
given by
 \be
h_{e}(x_{1},x_{2},b_{1},b_{2}) &=&S_{t}(x_{2})K_{0}(
\sqrt{x_{1}x_{2} }
m_{B}b_{1})  \nonumber \\
&&\times [ \theta (b_{1}-b_{2})K_{0}( \sqrt{x_{2}}
m_{B}b_{1}) I_{0}( \sqrt{x_{2} }m_{B}b_{2})  \nonumber
\\ && +\theta (b_{2}-b_{1})K_{0}( \sqrt{x_{2}}
m_{B}b_{2}) I_{0}( \sqrt{x_{2}} m_{B}b_{1}) ] \label{dh}\,,
\label{he} \\
 h_a(x_2,x_3,b_2,b_3)&=&\left(i\frac{\pi}{2}\right)^2
 S_{t}(x_{2})H_0^{(1)}(\sqrt{x_2x_3m_B^2}b_2) \non \\
 &\times & \left [\theta(b_2-b_3)H_0^{(1)}(\sqrt{x_3m_B^2}b_2)J_0(\sqrt{x_3m_B^2}b_3) \right.\non \\
 &+& \left. \theta(b_3-b_2)H_0^{(1)}(\sqrt{x_3m_B^2}b_3)J_0(\sqrt{x_3m_B^2}b_2) \right] \label{ha}\,,
 \ed
 and the evolution factor $E_{e, a}$ are defined by
 \be
E_{e}\left( t\right) &=&\alpha _{s}\left( t\right) S_{B}\left(
t\right)S_{P}\left( t\right)\;,
\nonumber \\
E_{a}\left( t\right) &=&\alpha _{s}\left( t\right) S_{P_1}(t)
S_{P_2}(t) \,, \label{Eea}
 \ed
where $S_{M}(t)$ denotes the Sudakov factor of M-meson and  $S_t$ is the threshold resummation effect. Their explicit expressions could be found in Ref. \cite{Chen:2002pz} and the
references therein. The hard scales for emission and annihilation are chosen to be
  \be
 t^1_e &=&\max ( \sqrt{x_{2} m_{B}^{2} } ,1/b_{1},1/b_{2}) \,,  \non\\
 t^2_e &=&\max ( \sqrt{ x_{1}m_{B}^{2}},1/b_{1},1/b_{2})\,,  \non \\
t^1_a &=&\max (\sqrt{x_3 m_B^2},1/b_2,1/b_3)\,,\non\\
t^2_a & =&\max (\sqrt{x_2 m_B^2},1/b_2,1/b_3).
 \ed

\end{appendix}


\begin{thebibliography}{99}

\bibitem{D0_PRL100} V.~M.~Abazov {\it et al.}  [D0 Collaboration],
  Phys.\ Rev.\ Lett.\  {\bf 100}, 142002 (2008)
  [arXiv:0712.0851 [hep-ex]].

\bibitem{CDF_PRL101}
 T.~Aaltonen {\it et al.}  [CDF Collaboration],
  Phys.\ Rev.\ Lett.\  {\bf 101}, 202001 (2008)
  [arXiv:0806.2472 [hep-ex]].

\bibitem{Aaltonen:2011kc}
  T.~Aaltonen {\it et al.}  [CDF Collaboration],
  arXiv:1101.0034 [hep-ex].

\bibitem{Campbell:1999ah}
  J.~M.~Campbell and R.~K.~Ellis,
  Phys.\ Rev.\  D {\bf 60}, 113006 (1999)
  [arXiv:hep-ph/9905386].


\bibitem{AKR}  O.~Antunano, J.~H.~Kuhn and G.~Rodrigo,
  Phys.\ Rev.\  D {\bf 77}, 014003 (2008)
  [arXiv:0709.1652 [hep-ph]].




\bibitem{Djouadi:2009nb}
  A.~Djouadi, G.~Moreau, F.~Richard and R.~K.~Singh,
  arXiv:0906.0604 [hep-ph].

\bibitem{Ferrario:2009bz}
  P.~Ferrario and G.~Rodrigo,
  Phys.\ Rev.\  D {\bf 80}, 051701 (2009)
  [arXiv:0906.5541 [hep-ph]].

\bibitem{Jung:2009jz}
  S.~Jung, H.~Murayama, A.~Pierce and J.~D.~Wells,
  Phys.\ Rev.\  D {\bf 81}, 015004 (2010)
  [arXiv:0907.4112 [hep-ph]].


\bibitem{Cheung:2009ch}
  K.~Cheung, W.~Y.~Keung and T.~C.~Yuan,
  Phys.\ Lett.\  B {\bf 682}, 287 (2009)
  [arXiv:0908.2589 [hep-ph]].

\bibitem{Frampton:2009rk}
  P.~H.~Frampton, J.~Shu and K.~Wang,
  Phys.\ Lett.\  B {\bf 683}, 294 (2010)
  [arXiv:0911.2955 [hep-ph]].

\bibitem{Shu:2009xf}
  J.~Shu, T.~M.~P.~Tait and K.~Wang,
  Phys.\ Rev.\  D {\bf 81}, 034012 (2010)
  [arXiv:0911.3237 [hep-ph]].

\bibitem{Arhrib:2009hu}
  A.~Arhrib, R.~Benbrik and C.~H.~Chen,
  Phys.\ Rev.\  D {\bf 82}, 034034 (2010)
  [arXiv:0911.4875 [hep-ph]].

\bibitem{Ferrario:2009ee}
  P.~Ferrario and G.~Rodrigo,
  JHEP {\bf 1002}, 051 (2010)
  [arXiv:0912.0687 [hep-ph]].

 \bibitem{Dorsner:2009mq}
  I.~Dorsner, S.~Fajfer, J.~F.~Kamenik and N.~Kosnik,
  Phys.\ Rev.\  D {\bf 81}, 055009 (2010)
  [arXiv:0912.0972 [hep-ph]].

\bibitem{Jung:2009pi}
  D.~W.~Jung, P.~Ko, J.~S.~Lee and S.~h.~Nam,
  Phys.\ Lett.\  B {\bf 691}, 238 (2010)
  [arXiv:0912.1105 [hep-ph]].

 \bibitem{Cao:2009uz}
  J.~Cao, Z.~Heng, L.~Wu and J.~M.~Yang,
  Phys.\ Rev.\  D {\bf 81}, 014016 (2010)
  [arXiv:0912.1447 [hep-ph]].


\bibitem{Cao:2010zb}
  Q.~H.~Cao, D.~McKeen, J.~L.~Rosner, G.~Shaughnessy and C.~E.~M.~Wagner,
  Phys.\ Rev.\  D {\bf 81}, 114004 (2010)
  [arXiv:1003.3461 [hep-ph]].

\bibitem{Kumar:2010vx}
  K.~Kumar, W.~Shepherd, T.~M.~P.~Tait and R.~Vega-Morales,
  JHEP {\bf 1008}, 052 (2010)
  [arXiv:1004.4895 [hep-ph]].

\bibitem{Martynov:2010ed}
  M.~V.~Martynov and A.~D.~Smirnov,
  arXiv:1006.4246 [hep-ph].

\bibitem{Chivukula:2010fk}
  R.~S.~Chivukula, E.~H.~Simmons and C.~P.~Yuan,
  arXiv:1007.0260 [hep-ph].

\bibitem{Bauer:2010iq}
  M.~Bauer, F.~Goertz, U.~Haisch, T.~Pfoh and S.~Westhoff,
  arXiv:1008.0742 [hep-ph].

\bibitem{Chen:2010hm}
  C.~H.~Chen, G.~Cvetic and C.~S.~Kim,
  Phys.\ Lett.\  B {\bf 694}, 393 (2011)
  [arXiv:1009.4165 [hep-ph]].

\bibitem{Jung:2010yn}
  D.~W.~Jung, P.~Ko and J.~S.~Lee,
  arXiv:1011.5976 [hep-ph].

\bibitem{Alvarez:2010js}
  E.~Alvarez, L.~Da Rold and A.~Szynkman,
  arXiv:1011.6557 [hep-ph].

\bibitem{Choudhury:2010cd}
  D.~Choudhury, R.~M.~Godbole, S.~D.~Rindani and P.~Saha,
  arXiv:1012.4750 [hep-ph].

\bibitem{Cheung:2011qa}
  K.~Cheung and T.~C.~Yuan,
  arXiv:1101.1445 [hep-ph].

\bibitem{Xiao:2011kp}
  B.~Xiao, Y.~K.~Wang, Z.~Q.~Zhou and S.~h.~Zhu,
  Phys.\ Rev.\  D {\bf 83}, 057503 (2011)
  [arXiv:1101.2507 [hep-ph]].

\bibitem{Delaunay:2011vv}
  C.~Delaunay, O.~Gedalia, S.~J.~Lee, G.~Perez and E.~Ponton,
  arXiv:1101.2902 [hep-ph].

\bibitem{Berger:2011ua}
  E.~L.~Berger, Q.~H.~Cao, C.~R.~Chen, C.~S.~Li and H.~Zhang,
  arXiv:1101.5625 [hep-ph].
%
\bibitem{Barger:2010mw}
  V.~Barger, W.~Y.~Keung and C.~T.~Yu,
  Phys.\ Rev.\  D {\bf 81}, 113009 (2010)
  [arXiv:1002.1048 [hep-ph]].

\bibitem{Amsler:2008zzb}
  C.~Amsler {\it et al.}  [Particle Data Group],
  Phys.\ Lett.\  B {\bf 667}, 1 (2008).

\bibitem{Barger:2011ih}
  V.~Barger, W.~Y.~Keung and C.~T.~Yu,
  Phys.\ Lett.\  B {\bf 698}, 243 (2011)
  [arXiv:1102.0279 [hep-ph]].

\bibitem{Bhattacherjee:2011nr}
  B.~Bhattacherjee, S.~S.~Biswal and D.~Ghosh,
  arXiv:1102.0545 [hep-ph].

\bibitem{Blum:2011up}
  K.~Blum {\it et al.},
  arXiv:1102.3133 [hep-ph].

\bibitem{Patel:2011eh}
  K.~M.~Patel and P.~Sharma,
  arXiv:1102.4736 [hep-ph].

\bibitem{Isidori:2011dp}
  G.~Isidori and J.~F.~Kamenik,
  arXiv:1103.0016 [hep-ph].

\bibitem{Foot:2011xu}
  R.~Foot,
  arXiv:1103.1940 [hep-ph].

\bibitem{Delaunay:2011gv}
  C.~Delaunay, O.~Gedalia, Y.~Hochberg, G.~Perez and Y.~Soreq,
  arXiv:1103.2297 [hep-ph].

\bibitem{AguilarSaavedra:2011vw}
  J.~A.~Aguilar-Saavedra and M.~Perez-Victoria,
  arXiv:1103.2765 [hep-ph].

\bibitem{Gresham:2011pa}
  M.~I.~Gresham, I.~W.~Kim and K.~M.~Zurek,
  arXiv:1103.3501 [hep-ph].

\bibitem{Shu:2011au}
  J.~Shu, K.~Wang and G.~Zhu,
  arXiv:1104.0083 [hep-ph].

\bibitem{Aaltonen:2011mk}
  T.~Aaltonen {\it et al.}  [CDF Collaboration],
  Phys.\ Rev.\ Lett.\  {\bf 106}, 171801 (2011)
  [arXiv:1104.0699 [hep-ex]].



\bibitem{Buchalla:1995vs}
  G.~Buchalla, A.~J.~Buras and M.~E.~Lautenbacher,
  Rev.\ Mod.\ Phys.\  {\bf 68}, 1125 (1996)
  [arXiv:hep-ph/9512380].

\bibitem{KpPQCD} H.N. Li and G. Sterman, Nucl. Phys. B{\bf 381}, 129 (1992); G.
Sterman, Phys. Lett. B{\bf 179}, 281 (1986); Nucl. Phys. B{\bf 281},
310 (1987); S. Catani and L. Trentadue, Nucl. Phys. B{\bf 327}, 323
(1989); Nucl. Phys. B{\bf 353}, 183 (1991); H.N. Li, Phys. Rev.
D{\bf 64}, 014019 (2001); H.N. Li, Phys. Rev. D{\bf 66}, 094010
(2002).


\bibitem{Chen:2002pz}
  C.~H.~Chen, Y.~Y.~Keum and H.~n.~Li,
  Phys.\ Rev.\  D {\bf 66}, 054013 (2002)
  [arXiv:hep-ph/0204166].

\bibitem{Chen:2008ie}
  C.~H.~Chen, C.~S.~Kim and Y.~W.~Yoon,
  Phys.\ Lett.\  B {\bf 671}, 250 (2009)
  [arXiv:0801.0895 [hep-ph]].


\bibitem{Li:2005kt}
  H.~n.~Li, S.~Mishima and A.~I.~Sanda,
  Phys.\ Rev.\  D {\bf 72}, 114005 (2005)
  [arXiv:hep-ph/0508041].

 \bibitem{TheHeavyFlavorAveragingGroup:2010qj}
  The Heavy Flavor Averaging Group {\it et al.},
  arXiv:1010.1589 [hep-ex].

\bibitem{YS1975} H.~C. Yen and I.~F. Shih, Chin. J. Phys.{\bf 13}, 205 (1975).


\bibitem{Lu:2002ny}
  C.~D.~Lu and M.~Z.~Yang,
  Eur.\ Phys.\ J.\  C {\bf 28}, 515 (2003)
  [arXiv:hep-ph/0212373].

\end{thebibliography}
\end{document}